\documentclass{aa}
\usepackage{graphicx}
\usepackage{natbib}

\bibpunct{(}{)}{;}{a}{}{,}

\begin{document} 

\title{The Sources of the CDF-N : broadband spectral analyses, absorption measurements and the $\alpha_{OX}-L_{UV}$ relation}

\author{S. Frank\inst{1}
           \and P. Osmer\inst{1}
           \and S. Mathur\inst{1}}

\offprints{S. Frank}
%\email{e-mail addresse}
\institute{Department of Astronomy, Ohio State University, 140 W.18th Ave., 
Columbus, OH 43210, USA}

\date{Received ??? / Accepted ???}
\abstract{We present broadband properties of sources in the Chandra Deep
Field-North (CDF-N) with known spectroscopic redshifts. The high
luminosity and high redshift bins are dominated by typical quasars. The
intermediate redshift (z $\sim 0.7-1.5$) and luminosity ranges
(L$_{0.5-8.0keV} \sim 10^{42.5-43.5}${} erg s$^{-1}$) show a mix of
different source types, with the apparently absorbed objects in the
majority. At the faint flux limit of the CDF-N, a substantial fraction
of the detections can be identified as normal/star-forming galaxies at
low redshift. The AGNs in the sample can be subdivided into four classes
based upon their rest frame spectral energy distribution: luminous,
apparently unabsorbed QSOs; objects with reddened optical spectra and no
signs of X-ray absorption; apparently X-ray absorbed AGNs with no signs
of reddening in their optical spectra; and optically reddened sources
with X-ray spectra indicative of large amounts of obscuration. We argue
that the AGNs of higher luminosity tend to have a lower X-ray absorbing
column density, but the ratio of X-ray absorbed to unabsorbed AGNs
remains constant with redshift. We find that the relations between UV
and X-ray luminosities derived by Strateva et al. (2005) and Steffen et
al. (2006) only hold for bright sources, and break down when faint
objects in the sample are included. This is only partly owing to the
fact that the majority of the faint sources are absorbed; several faint
sources must also have intrinsically lower X-ray luminosity. A fit to
the NIR-optical-UV broadband SEDs of AGNs allows us to constrain the
absorption parameters independently from the X-ray analyses. We show
that the N$_H${} values derived by the X-ray and optical methods are not
correlated. This may be because the X-ray absorption and optical
attenuation do not originate at the same location, and/or the dust
properties responsible for the optical attenuation in AGNs are very
different from the locally known dust properties.
}
\titlerunning{The Sources of the CDF-N} 
\authorrunning{S. Frank et al.}
\maketitle

%%%%%%%%%%%%%%%%%%%%%%%%%%%%%%%%%%%%%%%%%%%%%%%%%%%
%%%%%%%%%%%%%%%%%%%%%%%%%%%%%%%%%%%%%%%%%%%%%%%%%%%

\section{Introduction}\label{introduction}
The majority of the XRB below 8 keV has now been resolved by deep \em{Chandra}{} and \em{XMM-Newton}{} surveys \citep[]{2002ApJ...566L...5C, 2003ApJ...588..696M,2004MNRAS.352L..28W}{} - and the remaining uncertainty might be attributed to cosmic variation \citep[]{2003AN....324..165G,2003ApJ...585L..85Y}{} and problems in instrumental cross-calibration, although progress has been made for the latter problem \citep[]{2006ApJ...645...95H}. \citet[]{2004AJ....128.2048B}{} estimate the source densities for AGNs to reach $\sim$7200 sources deg$^{-2}${} and 4600 sources deg$^{-2}${} in the soft (0.5-2.0 keV at a limit of $\sim 2.5 \times 10 ^{-17}${} erg cm$^{-2}${} s${-1}$) and the hard band (2.0-8.0 keV at a limit of $1.4 \times 10 ^{-16}${} erg cm$^{-2}${} s${-1}$) for the Chandra Deep Fields(CDFs). These densities are factors of about 10-20 higher than for the deepest optical spectroscopic surveys. At these low X-ray flux limits, it becomes increasingly important to distinguish between emission of AGN and star-forming or even normal galaxies. \citet[]{2004AJ....128.2048B}{} arrive at a source density for star-forming galaxies at the same flux limits of $\sim$ 1700 and $\sim$ 700 sources deg$^{-1}${} in the soft and hard band, respectively. Their classification scheme is based upon a variety of different properties, such as the intrinsic X-ray luminosity, the spectral shape in the X-ray bands, variability, radio properties and features of the optical spectra as well as the X-ray-to-optical flux ratios. In this context, it is interesting to note that a Bayesian selected sample of \citet[]{2004ApJ...607..721N}{} from the same underlying source distribution using similar classification criteria, identifies a subsample from the CDFs as normal/starburst galaxies that is rather different from the \citet[]{2004AJ....128.2048B}{} sample. This discrepancy can only be explained partly by the use of a slightly different catalog by the two groups for the CDF-South and the restriction of \citet[]{2004ApJ...607..721N}{} to objects with spectroscopic redshifts z$_{spec} < 1.2$. It is obvious, that the determination of the host contribution to the flux in all bands remains a critical issue, and needs to be put on solid footing.\\
From examining the number counts of AGNs in the CDFs as functions of intrinsic X-ray luminosity and estimated absorption column density $N_{H}$, \citet[]{2004AJ....128.2048B}{} assert that the soft band (0.5-8.0 keV) XRB flux density is dominated by sources with $L_{0.5-8.0 keV} > 10^{43.5}${} erg s$^{-1}${} and $N_{H} < 10^{22}${} cm$^{-2}$, whereas main contributors to the hard band (2.0-8.0 keV) XRB flux density are primarily AGNs with intermediate luminosities ($L_{0.5-8.0 keV} = 10^{42.5} - 10^{43.5}${} erg s$^{-1}$) and a broad range of absorbing column densities. Comparing their number counts with results derived using the X-ray luminosity functions (XLFs) of \citet[]{2003ApJ...598..886U}, who examined a large sample of sources from different X-ray surveys with higher flux limits than the CDFs to estimate selection effects, \citet[]{2004AJ....128.2048B}{} note that their estimates are systemically and significantly lower than \citeauthor[]{2003ApJ...598..886U}'s, especially at the faint end of the flux range. A key element of these derivations are the assumed column density distributions for the AGNs. \citet[]{2004ApJ...616..123T}{} use a luminous subset of the complete CDFs sample to arrive at an observed N$_H${} distribution that shows significantly more sources with a column density higher than $10^{23}${} cm$^{-2}${} when compared to the distribution used by \citet[]{2003ApJ...598..886U}.\\
Absorbed AGNs are expected to be present to account for the emission budget of the diffuse X-ray background (XRB), as predicted first by \citet[]{1989A&A...224L..21S}{} and later more precisely modelled by e.g. \citet[]{1994MNRAS.270L..17M,1995A&A...296....1C,1999A&A...347..424G}. In these models, they are required to obtain the correct slope for the hard XRB and the source counts at flux limits of about 10$^{-13}${} erg sec$^{-1}$cm$^{-2}${} where their contribution of the total emission amounts to about 30\%{} \citep{2001A&A...366..407G}. \citet[]{1989A&A...224L..21S} required the ratio of unabsorbed to absorbed sources to be roughly one, simply achieved in their model by distributing sharp low energy cutoffs evenly between 10 and 30 keV for half of their sources. More refined models such as \citet[]{1999A&A...347..424G}{} introduce the two ratios R$_S${} and R$_Q${} of absorbed to unabsorbed Seyfert galaxies and QSOs as model parameters which can seperately vary with redshift. From a sample of nearby Seyfert galaxies, \citet[]{1995ApJ...454...95M}{} derived a ratio of type 2 to type 1 Seyfert of about 4. This estimate, however, is based upon a purely optical classification scheme with the indicator being the presence or absence of broad emission lines, and indeed the estimated X-ray column density distribution of the \citet[]{1995ApJ...454...95M}{} sample of Seyfert 2 galaxies shows that, although about 75\%{} of the sources are strongly affected by absorption ($N_{H} > 10^{23}${} cm$^{-2}$), there are some Seyfert 2 galaxies whose X-ray flux is virtually unabsorbed \citep{1999ApJ...522..157R}. It is possible that the extended emission contributes to the soft X-ray flux observed in such sources \citep[]{ghosh2007}.\\
Does the ratio of absorbed to unabsorbed AGNs change with redshift and/or luminosity ? This is not only an interesting question regarding the evolution of the population as a whole and the underlying physical mechanisms, but is of crucial importance when trying to map out the accretion history of the universe. While the extragalactic X-ray background models require a substantial number of absorbed sources, the search for them has remained inconclusive so far - and it is even much less clear how these sources could be distributed over redshift and luminosity. Studying the population found in soft and 'medium' (as opposed to hard, i.e. $>$10 keV say) X-ray surveys, various groups find different results : \citet[]{2005AJ....129..578B}{} report on a decreasing fraction of type 1 AGNs with redshift (their fig.19), and the analysis of \citet[]{2006ApJ...639..740B}{} derives an evolving type 2 AGN fraction as $R \sim (1+z)^{0.3}$. Recently, La Franca et al. (2007), updating their results from \citet[]{2005ApJ...635..864L} see a strongly rising fraction of X-ray absorbed sources with redshift in the CDF-S when applying spectral analyses and correcting for detection probabilities. The analysis of \citet[]{astro-ph0610939}, however, including the largest well characterised X-ray selected sample to date shows no redshift dependence of the absorbed fraction of quasars. Such an evolution is also not needed by the models of e.g. \citet[]{2003ApJ...598..886U}{} and \citet[]{2004ApJ...616..123T}. Interestingly, the latter two groups differ in the requirement of a luminosity dependence of this fraction : while Treister et al. operate with an underlying model that does not change with luminosity and thus attributes all of the apparent evolution to observational effects, Ueda et al. favour a scenario where the more luminous quasars tend to be less X-ray absorbed. \citet[]{astro-ph0610939}{} also find evidence for a decreasing fraction of absorbed sources with luminosity. \citet[]{2006ApJS..165...19E}, however, analysing data from the SEXSI survey find a fraction of absorbed sources that does not even evolve with source luminosity (their fig. 22). Given all these at first hand contradictory results, we revisit this question.\\
The relation between the X-ray and UV luminosities of quasars, usually examined via the slope of a hypothetical power-law connecting the monochromatic luminosities at rest-frame frequencies $\nu$, defined by \citet[]{1979ApJ...234L...9T}, $\alpha _{OX} = log (l_{2keV}/l_{2500 \AA}){} /{} log (\nu_{2 keV}/\nu_{2500\AA})$, has been the object of a variety of studies \citep[]{1982ApJ...262L..17A, 1985ApJ...297..177K, 1986ApJ...305...83A, 1987ApJ...314..111A}. Recently, \citet[]{2005AJ....130..387S} have used an optically selected sample of quasars across a wide redshift range with ROSAT detections to confirm and improve the anti-correlation of $\alpha _{OX}${} with the UV luminosity generally found in earlier studies \citep[]{1994ApJS...92...53W, 2003AJ....125..433V}. \citet[]{2006AJ....131.2826S} have substantially extended these analyses by including moderate-luminosity, optically selected AGNs from the COMBO-17 survey with corresponding deep X-ray observations from the Extended $Chandra${} Deep Field-South, and luminous AGNs from the Bright Quasar Survey. Extensive partial-correlation studies of their full sample of 333 AGNs reveal no significant correlation of $\alpha_{OX}${} with redshift, but a strong dependence of $\alpha_{OX}${} on UV luminosity. These empirical relations between intrinsic quantities of the AGNs are clearly important for our understanding and testing models of their energy generation mechanism, and bolometric correction estimates. Furthermore, \citet[]{2006AJ....131.2826S} stress that a comparison of luminosity functions derived in different bands and at different epochs crucially hinge upon the evolution of $\alpha_{OX}$. While the majority of these studies have relied on optically selected samples, and - in order to study intrinsic AGN properties - carefully pruned their object lists to exclude absorbed sources, it is an interesting question to ask whether these relations hold for X-ray selected AGNs and - if not - whether absorption alone can explain the discrepancy.\\   
In the following work, we explore properties of the sources in Chandra Deep Fields (CDF), first by presenting their rest frame broadband spectral energy distributions (SED). We also derive an estimate of the observed column density distribution of the complete sample of sources with known spectroscopic redshifts, using the ratio of soft to hard band count rates. Both redshift and luminosity effects are responsible for an evolution of this distribution, and we try to assess the relative importance of both effects. Finally, in order to be able to compare results of optical/UV AGN surveys with the CDFs, we test whether correlations that link the X-ray with the rest-frame UV properties of AGNs hold for this very faint sample.\\ 
The organisation of the paper is as follows : in section \ref{Data_analysis}{} we give an overview of the data sets used. In section \ref{Source_classification}, we present the broadband SEDs and the classification of the sources. Section \ref{NHestimate}{} presents the estimate for the X-ray absorbing hydrogen column density N$_H${} and a discussion of the different evolutionary and observational effects of redshift and luminosity on the distribution of N$_{H}$. The derivation of an optical N$_{H}${} and comparisons to the X-ray derived column densities are presented in that section as well. The correlations of \citet[]{2005AJ....130..387S}{} and \citet[]{2006AJ....131.2826S}{} are tested in section \ref{UV-Xray}. We summarise and discuss our findings in section \ref{Discussion}, where we also give an outlook to future steps following our analysis.\\ 
Throughout this paper, we adopt the cosmological model with $H_{0} = 71${} km s$^{-1}${} Mpc$^{-1}$, $\Omega _{M} = 0.27${} and $\Omega _{\Lambda} = 0.73$. 
%%%%%%%%%%%%%%%%%%%%%%%%%%%%%%%%%%%%%%%%%%%%%%%%%%%
%%%%%%%%%%%%%%%%%%%%%%%%%%%%%%%%%%%%%%%%%%%%%%%%%%%

\section{Data analysis}\label{Data_analysis}
Our primary goal is to analyse information from publicly available data sets with an emphasis on the emitted-frame picture. The following sections list the samples we utilised.

%%%%%%%%%%%%%%%%%%%%%%%%%%%%%%%%%%%%%%%%%%%%%%%%%%%

\subsection{X-ray Samples}
We use data extracted from the \citet[]{2003AJ....126..539A} point-source catalogue for the $\sim$2 Ms exposure of the Chandra Deep Field North. In their data tables 3a and 3b, 503 point-sources are listed with their positions in the 448 arcmin$^2$ field, count rates in 7 standard bands, the effective exposure times, estimates for the effective photon index $\Gamma _{eff}${} for a power-law model with the Galactic column density and observed-frame fluxes in the seven bands. Additionally, we incorporate the X-ray data for the 1 Ms CDF-S from the catalogue of \citet[]{2002ApJS..139..369G} which lists data for 346 point-sources.

%%%%%%%%%%%%%%%%%%%%%%%%%%%%%%%%%%%%%%%%%%%%%%%%%%%%

\subsection{Additional data}
Both of the deep X-ray surveys were accompanied by extensive follow-up observations in a variety of bands. Here we mainly use the NIR-optical-UV data described in \citet[]{2003AJ....126..632B}{} for the CDF-N and the catalogue of photometric redshifts of 342 sources for the CDF-S by \citet[]{2004ApJS..155...73Z}. The former contains broadband photometric measurements in U (obtained with MOSAIC at the Kitt Peak 4m Telescope), B, V, R, i, z' (Suprime-cam at the 8.2m Subaru Telescope) and HK' (QUIRC at the 2.2m University of Hawaii Telescope) and spectroscopic redshifts based upon observations with DEIMOS at KeckII for 270 sources. The latter assigns redshifts estimates to 342 of the 346 CDF-S sources, an impressive 98.9\%{} completeness. 173 among these have also reliable spectroscopic determinations of their redshifts, which can be used to gauge the precision of the photometric estimates.  

%%%%%%%%%%%%%%%%%%%%%%%%%%%%%%%%%%%%%%%%%%%%%%%%%%%%%%%%%%
%%%%%%%%%%%%%%%%%%%%%%%%%%%%%%%%%%%%%%%%%%%%%%%%%%%%%%%%%%

\section{Broadband Spectral Energy Distributions}\label{Source_classification}
\begin{figure}
\resizebox{\hsize}{!}{\includegraphics[angle=270]{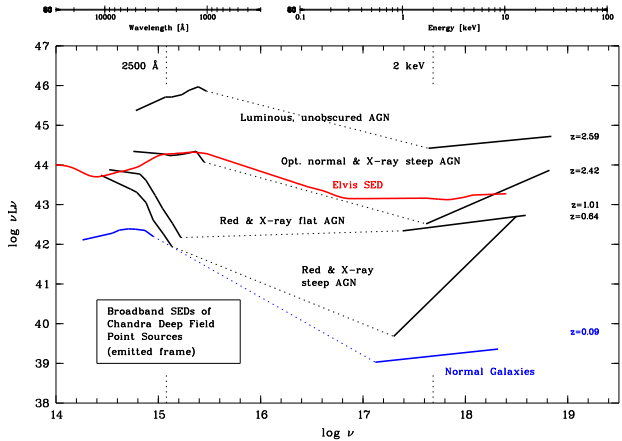}}
\caption[Broadband SEDs of 5 typical CDF-N sources]{Broadband Spectral Energy Distributions of 5 typical CDF-N point sources in the emitted frame. The source redshift is denoted to the right. Photometric data are shifted from the observed HK',z',I, R, V, B and U bands to the emitted frame. The X-ray part of the spectrum is assumed to be a power-law with an observed effective photon index $\Gamma _{eff}$. Overplotted in red is the standard AGN SED of \citet{1994ApJS...95....1E}.}
\label{main_plot_4different_types}
\end{figure}

On the L-z plane, the sources of the CDFs cover a wide range in both luminosity (39.0 $\leq$ log L$_{X, 0.5-8.0 keV}${} $\leq$45.0) and redshift (0.08 $\leq$ z$_{spec} \leq$ 5.3). Most of the earlier studies have focused on the observed properties of the objects. It is, however, important to know their intrinsic differences and similarities. Plotting the sources in the emitted frame allows for a direct visualisation of the dramatic differences in luminosity and redshifts that characterise the different source types. Furthermore, the effects of absorption on both the X-ray and the NIR-optical-UV parts of the spectra become immediately evident. Fig.~\ref{main_plot_4different_types}{} shows the broadband spectral energy distributions of four typical sources in the Chandra Deep Field-N survey. In the X-rays, the objects span a luminosity range of over 6 orders of magnitudes, while the range of their NIR-optical-UV luminosities is still an impressive 4 orders of magnitude. The observed NIR and optical bands shift completely to emitted UV for higher redshifts, whereas the observed X-ray regime from 0.5-8.0 keV transforms into the hard X-ray bands from $\sim$3-50 keV for the sources of the highest redshift ($z\sim5.0$). There are large variations both in the X-ray slopes and the optical broadband spectral energy distributions amongst the various sources - and thus significant deviations from the standard SED of luminous AGN compiled by \citet{1994ApJS...95....1E}.\\

The deep 2Msec exposure probes very faint X-ray sources hitherto not studied at cosmological distances. In fact, using Bayesian statistical methods, using the X-ray luminosities, X-ray hardness ratios, and the X-ray to optical flux ratios as discriminators, \citet[]{2004ApJ...607..721N}{} classify 129 of 270 objects in the CDF-N with known spectroscopic redshifts as normally starforming or starburst galaxies with no or little AGN contribution. An example of such a galaxy's broadband SED is shown in blue in Fig. \ref{main_plot_4different_types}. We caution, however, that \citet[]{2004AJ....128.2048B}{} identify a different subset of the same sources as galaxies, using two slightly different classification schemes that also take radio morphologies, variabilities and optical spectroscopic classifications into account. Even their {\em optimistic}{} galaxy sample is much smaller than the \citet[]{2004ApJ...607..721N}{} sample, and they expect significant AGN contamination present in the latter. For a complete discussion see \citet[]{2004AJ....128.2048B}.\\
We grouped the remaining AGNs simply by their visual appearance in the log $\nu$L$_{\nu}${} - log $\nu${} -plot into four groups :
\begin{itemize}
\item Luminous, apparently unabsorbed AGN
  \begin{itemize}
  \item The X-ray power-law slope is close to the canonical $\Gamma _{eff} = 1.9${} of the \citet{1994ApJS...95....1E}{} AGN sample.
  \item The optical/UV data show a rise or stay flat shortward of  2500 \AA.
  \item The optical-to-X-ray flux ratio is close to the \citet{1994ApJS...95....1E}{} value.
  \end{itemize}
\item Red \&{} X-ray flat AGN\footnote{Here, 'flat' refers to the SED in the log $\nu L_{\nu}${} - log $\nu${} -plot, i.e. a $\Gamma _{eff} \sim 2.0.$} 
  \begin{itemize}
  \item The X-ray power-law slope is close to the canonical $\Gamma _{eff} = 1.9${} of the \citet{1994ApJS...95....1E}{} AGN sample. 
  \item The optical/UV data show a sharp decline shortward of 2500 \AA, indicative of extinction.
  \end{itemize}
\item Red \&{} X-ray steep AGN
  \begin{itemize}
  \item The X-ray power-law slope is much less than the canonical $\Gamma _{eff} = 1.9${} of the \citet{1994ApJS...95....1E}{} AGN sample, indicative of either an intrinsically hard spectrum or absorption. 
  \item The optical/UV data show a sharp decline shortward of 2500 \AA.
  \end{itemize}
\item Optically normal \&{} X-ray steep AGN
  \begin{itemize}
  \item The X-ray power-law slope is much less than the canonical $\Gamma _{eff} = 1.9${} of the \citet{1994ApJS...95....1E}{} AGN sample, indicative of either an intrinsically hard spectrum or absorption. 
  \item The broadband SED in the NIR-optical-UV does not show signs of signficant absorption, i.e. the optical/UV data show a rise or stay flat shortward of 2500 \AA.  \end{itemize}
\end{itemize}

\begin{figure}
\resizebox{\hsize}{!}{\includegraphics[angle=270]{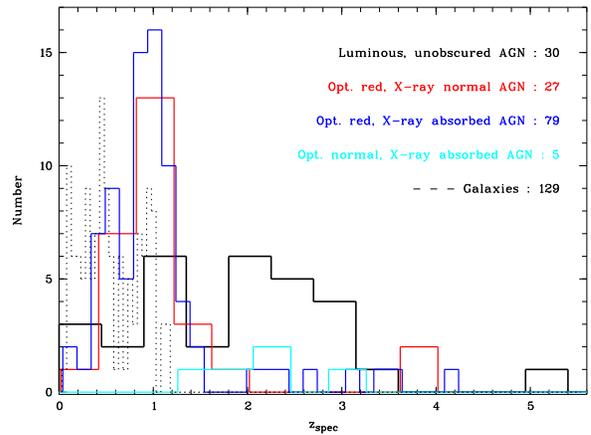}}
\caption[Redshift distribution of the 5 classes of sources.]{Redshift distribution of the 5 classes of objects in the CDF-N. Note that the classification scheme of \citet{2004ApJ...607..721N}{} does not allow for galaxies to have redshift higher than 1.2. The fraction of AGNs exhibiting signs of obscuration in the optical/UV and/or the X-rays, based upon the visual classification scheme proposed in the text, changes rapidly with redshift. Note the reduced bin sizes for the red \&{} X-ray steep AGNs and the galaxies.}
\label{redshift_classification}
\end{figure} 
The redshift distribution of these 5 classes of objects is illustrated in Fig.~\ref{redshift_classification}. Not surprisingly, the lowest redshift bins, presumably with the lowest luminosity sources, are dominated by the galaxies, whereas the highest redshift objects tend to be luminous, unobscured AGNs. The intermediate redshift bins for the AGN group of sources, however, show a mix of the different types, with the apparently unobscured objects in the minority. Note that the classification scheme of \citet{2004ApJ...607..721N}{} only allows for galaxies to have redshifts z$<$1.2.\\
In Appendix A, the sources are listed as belonging to one of the above classes. In addition, we include all the information derived in the following paragraphs for each source (absorbing X-ray column density N$_{H}$, rest-frame luminosities observed and absorption corrected, NIR-UV fits to the dust extinction, etc...). 

%%%%%%%%%%%%%%%%%%%%%%%%%%%%%%%%%%%%%%%%%%%%%%%%%%%%%%%%%%%%
%%%%%%%%%%%%%%%%%%%%%%%%%%%%%%%%%%%%%%%%%%%%%%%%%%%%%%%%%%%%

\section{Estimating the X-ray absorbing hydrogen column density N$_{H}$}\label{NHestimate}
The observed count rate in the X-rays can be strongly affected by the presence of absorbing gas in the sightline towards the AGN, with the soft X-ray bands diminished much more strongly. The ratio of count rates in the soft (0.5-2.0 keV) to the hard X-ray band (2.0-8.0 keV) can thus be used to estimate the column density of the intervening gas, provided the originally expected ratio is known. 

\begin{figure}
\resizebox{\hsize}{!}{\includegraphics[angle=270]{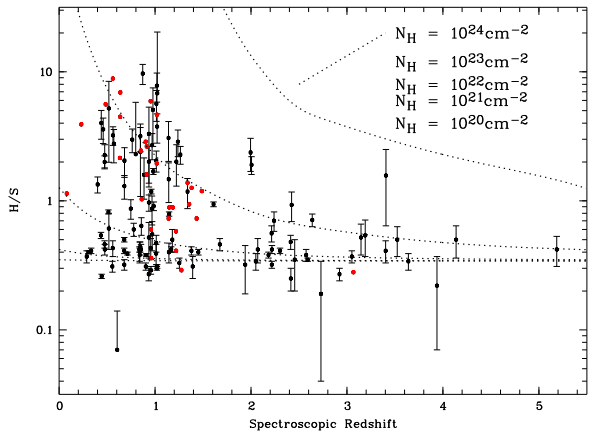}}
\caption[Hard-to-soft-band count rate.]{The ratio of counts in the hard (2.0 - 8.0 keV) to the soft (0.5-2.0) band of the CDF-N sources classified as AGNs versus their spectroscopically confirmed redshift. The sources marked in red do not have reliable (or sensible) error estimates due to the low count rate in one or both bands. The dotted lines give the expected H/S for a simple power-law ($\Gamma = 1.8$) model for various absorbing column densities $N _{H}$. Note the extremely narrowing range for H/S with increasing redshift - at redshifts $z > 3.0${} it is difficult to reliably estimate an $N_{H} < 10^{23}cm^{2}${} with this method.}
\label{zspec_HS}
\end{figure}

Fig. \ref{zspec_HS}{} shows the distribution of the hard-to-soft-band count rate $R = H/S$ of the CDF-N AGN sample sources versus their redshifts. We have calculated the ratios expected for a simple power-law model of the original spectrum and various column densities N$_H$. The photon index $\Gamma$ of the original spectrum is assumed to be 1.8, in agreement with the findings of \citet[]{2006A&A...451..457T}. We have performed all of the following analyses with different photon indices (ranging from 1.5 to 2.2), with all the major results still holding. The calulations were performed with PIMMS\footnote{PIMMS is the {\em P}ortable{\em I}nteractive{\em M}ulti-{\em M}ission{\em}Simulator, originally written by Koji Mukai. For further information see http://heasarc.gsfc.nasa.gov/docs/software/tools/pimms.html.}, using the observed galactic N$_{H}${} column density $(1.6 \pm 0.4) \times{} 10^{20}${}cm$^{-2}$, \citet[]{1992ApJS...79...77S}) and the appropriate Chandra Observing cycle. Note that PIMMS only takes into account photoelectric absorption processes. R = R(N$_{H}$,z) is a strong function of the absorbing column density and the redshift.  From Fig. \ref{zspec_HS}{} it is evident that even small errors in the count rate ratios can introduce large over- or underestimates of absorbing column densities. Furthermore, the strong non-linearity of the relation between count rate ratio and column density can introduce a significant bias towards lower column density values even for error distrubutions that are symmetric around the mean H/S ratio. These effects are especially pronounced for the high redshift sources. In order to assess the severity of this bias on the observed column density distribution, we have performed a Monte-Carlo-simulation with 10,000 representations of the sample, in which we randomised each source's ratio according to its individual error while keeping its redshift fixed. These representations allow us not only to estimate the aformentioned effects on the distribution as a whole, but also provide us with an error estimate for the column density for each single source. While individual estimates of N$_H${} with the simple hardness ratio method remain fairly insecure (the average spread in log N$_H${} is 0.6 dex), the overall shape of the complete distribution is very robust against the randomisation process, proving the usefulness of the simple estimation method on statistically large enough samples.  
\begin{figure}
\resizebox{\hsize}{!}{\includegraphics[angle=270]{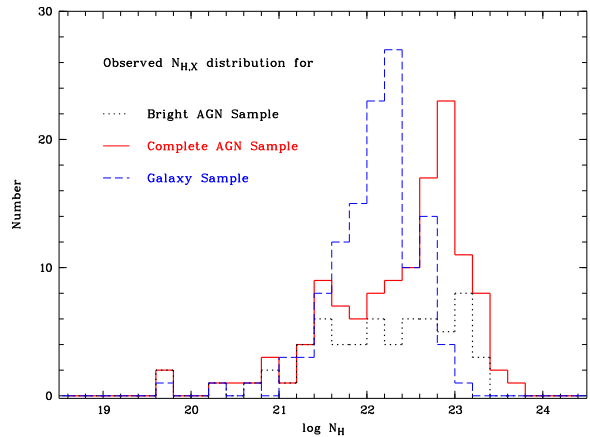}}
\caption[Observed N$_{H}${} column density distribution.]{The observed column density distribution for the 270 CDF-N sources with reliable spectroscopic redshifts. The distribution is derived by averaging over 10,000 representations of the sample's band rate ratios as in Fig.\ref{zspec_HS}{}, randomised according to their errors. The objects classified as galaxies have in general a much lower column density (under the specific assumptions for the X-ray spectrum as detailed in the text). Note how the peak of the AGN distribution shifts to a higher column density when including the X-ray faint objects.}
\label{obs_NH_distribution_different_types}
\end{figure}

\begin{figure}
\resizebox{\hsize}{!}{\includegraphics[angle=270]{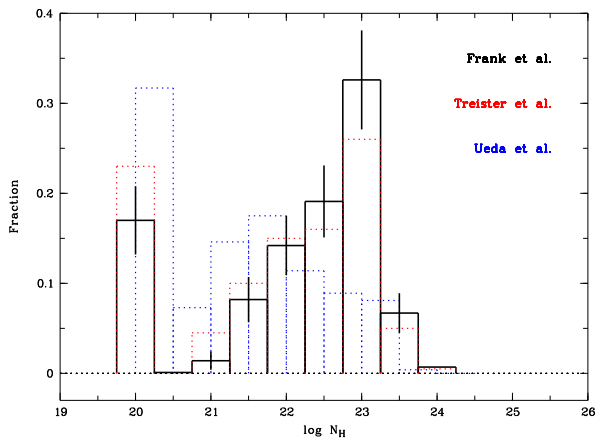}}
\caption[Observed N$_{H}${} column density distributions.]{Observed fractional column density distributions for AGNs of various analyses. The distribution for the CDF-N sources classified as AGN derived in this work is the black line and has errors estimated from the range of values in the different bins from the 10,000 representations of the Monte-Carlo simulation. The dotted red histogram represents the distribution derived by \citet{2004ApJ...616..123T}{} which is based upon the two GOODS Fields N and S. The histogram in blue is the distribution used by \cite[]{2003ApJ...598..886U}, which comprises a selection of sources from samples with higher soft X-ray flux limit. The lowest N$_H${} bin contains all sources whose column density could not be constrained well by the method presented in the text.} 
\label{obs_NH_distribution_different_studies}
\end{figure}

Fig. \ref{obs_NH_distribution_different_types}{} shows the observed overall distribution of column density derived by averaging over the random data sets, subdivided into the AGN and galaxy samples. Sources with unconstrained N$_H${} were left out of this plot; we included them, however, in the analysis of the fractional distribution (cf. plot \ref{obs_NH_distribution_different_studies}). The galaxies have in general a lower column density, however we caution that this could purely be an artefact of the simple assumptions used to model the X-ray spectra, which in the case of normal or star-forming galaxies certainly originates from a variety of sources (cf. e.g. the review by \citet[]{1989ARA&A..27...87F}) and thus are not necessarily  well represented by simple power-law.\\
In stark contrast to \citet[]{2003ApJ...598..886U}{}, who operate with an observed column density distribution that peaks around log $N_{H} \sim 21.5${} cm$^{-2}${} for their overall sample (cf. Fig. 5 of their analysis), we obtain a distribution that still sees a strong rise in the number of observed sources up to a column density of $10^{23}${} cm$^{-2}$. Our distribution for the CDF-N is very similar to the results of \citet[]{2004ApJ...616..123T}, who derived intrinsic obscuring column densities for the two GOODS fields' objects with spectroscopic redshift in a similar procedure to ours. Fig. \ref{obs_NH_distribution_different_studies}{} clearly demonstrates the similiarity of our results with those of \citet{2004ApJ...616..123T}{} while also highlighting the difference between these two distributions and the one used by \citet{2003ApJ...598..886U}. Applying the same bright flux cut-off of $3.0 \times 10^{-15}${} erg s$^{-1}${} cm$^{-2}${} (for the full 0.5 - 8.0 keV Chandra band) as \citet[]{2003ApJ...598..886U}{} to the AGN sample, results primarily in a loss of sources with higher column densities and thus a dramatic difference in  the shape of the column density distribution (cf. fig. \ref{obs_NH_distribution_different_types}).\\
Since we are using an almost identical approach as \citet{2004ApJ...616..123T}{} in estimating the column density N$_H$ and furthermore analyse a subsample of the CDF-N sources that comprises a good part of the \citet{2004ApJ...616..123T} sample, it is not surprising that the two distributions are very similar. We do note, however, that the identification of sources as AGNs in our case relies upon the Bayesian scheme of \citet{2004ApJ...607..721N}, whereas \citet{2004ApJ...616..123T}{} simply apply a luminosity cut (log L$_{X}$(2.0-8.0 keV) $> 42.0$) to a subsample of the GOODS field that includes only sources with unambigious optical counterparts to spectroscopically identified X-ray sources. Sampling the complete CDF-N region, which is slightly larger than the GOODS field, and applying the aforementioned classification scheme of \citet{2004ApJ...607..721N}{}, our AGN sample consists of 141 objects. 16 of these have a hardband luminosity of less than $10^{42}$erg s$^{-1}$, and an additional 18 of the remainder show an optical counterpart, which is offset from the X-ray centroid by more than 1.0''. If, however, we apply the simple cuts proposed by \citet{2004ApJ...616..123T}{} to the complete set of sources with spectroscopic redshifts, 10 sources that \citet{2004ApJ...607..721N}{} classify as galaxies remain in the resulting subsample. All but the two X-ray brightest of these 10 objects, however, exhibibit signs of strong X-ray obscuration (log N$_H > 22.0$), and thus their inclusion in the sample would not change the shape of the column density distribution much.\\ 
The column density distribution for the CDF-S field, which had to rely on less secure photometric redshifts, is not shown here, but shows essentially the same behaviour.\\
\citet[]{2004ApJ...616..123T}{} model the N$_H${} distribution by invoking an obscuring torus of randomly distributed, optically thick gas and dust clumps engulfing the central black hole. The geometry and gas-to-dust ratio of the torus are selected such that the ratio of obscured to unobscured sources is $\sim$3. As these parameters in the \citet[]{2004ApJ...616..123T}{} model do not change with redshift or intrinsic luminosity of the AGN, this ratio remains constant. The following paragraph presents our analysis as to whether these assumptions hold.   

%%%%%%%%%%%%%%%%%%%%%%%%%%%%%%%%%%%%%%%%%%%%%%%%%%%%%%%%%%%%%%%%%

\subsection{Redshift and luminosity effects on the column density distribution}
The presence of gas in the sightline not only affects the hardness ratio, but also removes the more absorbed sources from flux-limited samples. This effect is redshift dependent because sources at higher redshifts are observed at higher energies (in the emitted frame) and are thus less affected. Furthermore, the obscuration could be luminosity dependent as emission from the central source affects the state of the surrounding material. Thus, it is crucial to infer information about how the column density distribution changes with both redshift and luminosity.
 
\begin{figure}
\resizebox{\hsize}{!}{\includegraphics[angle=270]{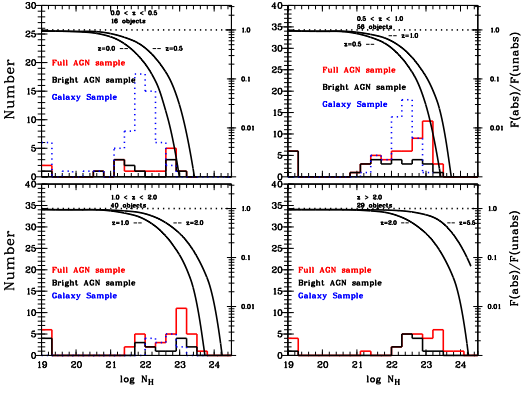}}
\caption[Observed N$_{H}${} column density distribution for z<0.5]{The observed column density distribution for CDF-N sources in different redshift ranges (left axes). The lines indicate the amount of absorption in the soft X-ray band, assuming an intrinsic power-law spectrum with photon index $\Gamma = 1.8${} and a source redshift as indicated. (right axes. A value of 0.1 represents 90\%{} flux absorption). Note how the peak of the AGN column density distribution for $z < 1.0${} coincides with the column density where approximately only 10\%{} of the emitted flux in the soft band is not absorbed, and how the shape of the distribution changes when including the faint sources. The number listed underneath the redshift range at the top refers to the number of AGNs in this bin.  The almost complete absence of galaxies in the bins above $z = 1.0${} results from the redshift criterion used by \citet{2004ApJ...607..721N}.}
\label{four_plots_diff_z}
\end{figure}
\begin{figure}
\resizebox{\hsize}{!}{\includegraphics[angle=270]{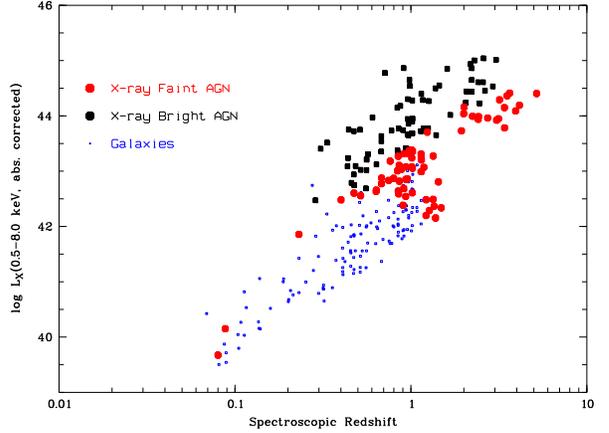}}
\caption[X-ray Luminosity vs. redshift]{The full-band X-ray luminosity of the 270 CDF-N sources versus their spectroscopically determined redshifts. The luminosities are absorption corrected, assuming an intrinsic power-law slope for each source with photon index $\Gamma = 1.8$. Note that the fainter sources are also less luminous than their bright counterparts at the same redshifts, even after correcting for absorption. The dearth of sources classified as AGNs from redshifts $\sim 1.3 - 2.2${} can be attributed to the difficulty of reliably determining redshifts in this interval due to the lack of useful emission lines in the spectra.}
\label{zspec_log_vLv}
\end{figure}

Figure \ref{four_plots_diff_z}{} shows the distribution of observed column densities broken down by 4 different redshift bins. The peak of the distribution shifts to higher values for increasing redshifts. For redshifts lower than roughly 1.5, the high N$_{H}${} cut-off, beyond which the number of sources rapidly dwindles, coincides with the value of N$_{H}${} for which of order 90\%{} of the soft-band flux is absorbed (right axes in fig. \ref{four_plots_diff_z}). Thus, the prevalence of sources with higher column densities at higher redshifts can not easily be attributed to just an evolutionary effect, but could rather represent an observational bias. Ironically, it becomes progressively easier with increasing redshifts to detect sources of high column densities as one samples increasingly harder parts of the emitted X-ray spectrum - note that the peak of the column density distribution at redshifts beyond $z \sim 1.5${} is not coincident any more with the position of $N_{H}${} necessary to absorb about 90\%{} of the soft-band flux.\\
Fig. \ref{obs_NH_distribution_different_types}{} shows that the X-ray faint AGNs of the CDF-N survey in general have higher absorbing column densities, and Fig. \ref{zspec_log_vLv}{} demonstrates that these are primarily sources of lower luminosities (log L$_{X} < 43.5$) than their counterparts at similar redshifts (z$\sim$0.5-1.1). Interestingly, this subsample of the sources is almost completely comprised of objects identified as red and X-ray steep AGNs in our visual classification scheme of section \ref{Source_classification}, i.e. they show signs of absorption in the optical, as well as in X-rays. Thus, fig.\ref{obs_NH_distribution_different_types}{} demonstrates that these less luminous sources are more absorbed.\\

\begin{figure}
\resizebox{\hsize}{!}{\includegraphics[angle=270]{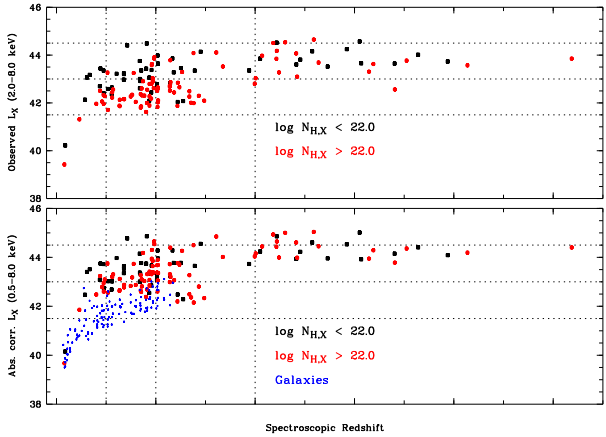}}
\caption[Redshift vs. Luminosity and colour coded absorption]{Upper panel : The observed rest-frame hard-band X-ray luminosity for the CDF-N AGNs versus their spectroscopically determined redshifts. Data points in red represent sources with log N$_H > 22.0$, indicative of heavy X-ray absorption. The dotted lines mark borders of the bins in redshift and luminosity to analyse evolutionary effects. Note the disjoint distribution at the lower redshift and luminosity bins.\\Lower panel : The absorption corrected full-band X-ray Luminosity for the CDF-N sources versus their spectroscopic redshifts. The data points in blue represent the objects classified by Norman et al. (2004) as galaxies. Notice how the absorption correction leads to a clearer disentanglement between those sources and the AGNs, as well as a more homogeneous distribution of the absorbed and unabsorbed AGNs in the low luminosity, low redshift regime. The dotted lines mark again the borders of the bins in redshift and luminosity to analyse evolutionary effects.}
\label{z_spec_vs_Lx_I}
\end{figure}

%\begin{figure}
%\resizebox{\hsize}{!}{\includegraphics[angle=270]{fraction_obs_unabs_vs_redshift.ps}}
%\caption[Redshift vs. fraction]{The fraction of sources with log N$_H >${} 22.0 for the AGN in the CDF-N versus redshift. The data points in black represent the complete sample, whereas the red and green values are derived taking into account only sources with an observed rest-frame fullband X-ray luminosity of log L$\_X$(0.5-8.0keV) below and above 43.0, respectively. Note that there no absorbed sources below a redshift of z$_{spec} <${} 0.5 for the latter, whereas there is only one such AGN above z$_{spec}${} = 2.0 for the former.}
%\label{fraction_vs_zspec_I}
%\end{figure}

%\begin{figure}
%\resizebox{\hsize}{!}{\includegraphics[angle=270]{z_spec_vs_unabs_Lx_abs_vs_unabs.ps}}
%\caption[Redshift vs. L_X unabsorbed]{Absorption corrected full-band X-ray luminosities versus spectroscopic redshifts for the CDF-N sample. Data points in red represent sources with log N$_H > 22.0$, indicative of heavy X-ray absorption. The dotted lines mark borders of the bins in redshift and luminosity to analyse evolutionary effects. The green data points are objects classified as galaxies.}
%\label{z_spec_vs_Lx_II}
%\end{figure}

\begin{figure}
\resizebox{\hsize}{!}{\includegraphics[angle=270]{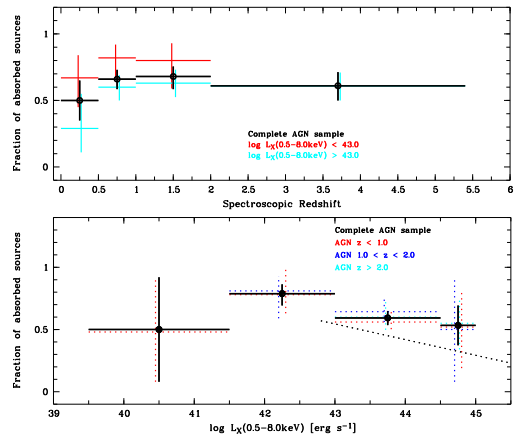}}
\caption[Redshift vs. fraction]{Upper panel : The fraction of sources with log N$_H > 22.0${} for the AGNs in the CDF-N versus redshift. The data points in black represent the complete sample, whereas the red and green values are derived taking into account only sources with an absorption-corrected rest-frame fullband X-ray luminosity of log L$_X$(0.5-8.0keV) below and above 43.0,respectively.\\Lower panel :The fraction of sources with log N$_H > 22.0${} for the AGNs in the CDF-N versus their absorption corrected luminosity. The data points in black represent the complete sample, whereas the red,green and blue values are derived taking into account only sources within the indicated redshift bins. The dotted line represents the \citet[]{2003ApJ...598..886U} fit to their data.}
\label{fraction_vs_zspec_II}
\end{figure}

Is the absorption redshift dependent ? In order to examine evolutionary effects, we have divided the sample of AGNs in bins of redshift and luminosity. Fig. \ref{z_spec_vs_Lx_I}{} shows the distribution of absorbed and unabsorbed sources in the observed hard-X-ray rest-frame luminosities versus the sources' redshift. AGN with an absorbing column density log N$_H > 22.0${} are denoted in this plot with red symbols. Clearly, the sources of the low redshift and luminosity bins are dominated by absorbed objects, as expected, whereas more luminous AGNs tend to be unabsorbed at redshifts below 1.5. The overall ratio of absorbed to unabsorbed sources, integrating over all luminosities present, remains constant with redshift, but it does not quite reach the \citet[]{1995ApJ...454...95M}{} value of R$\sim$0.8, as can be seen in fig. \ref{fraction_vs_zspec_II}. The lower fraction of absorbed sources at redshifts below $\sim$0.5 can be readily attributed to the effects of redshift-dependent removal of sources from a flux-limited sample as discussed above. When considering only sources with an observed full-band luminosity log L$_X$(0.5-8.0keV) $< 43.0$, about 85\% of AGNs exhibit an absorbing X-ray column density above N$_{H} = 10^{22}$cm$^{-2}$ for the redshift range $0.5 < z < 2.0${} - we note however, that this criterion of classifying an AGN as absorbed does not coincide with the original \citet[]{1995ApJ...454...95M}{} scheme that rests upon the existence or non-existence of broad emission lines in the optical spectra of Seyfert galaxies. Interestingly, the fraction of AGNs with observed X-ray luminosities above L$_{X}(0.5-8.0 keV) = 10^{43.0}${} erg s$^{-1}${} that exhibit strong absorption rises rapidly with redshift. This rather surprising result can be explained mainly by two factors. At low redshifts, the volume probed is too small to expect many sources of high luminosities : in order to be able to obtain enough flux in the soft-band to determine the column density (cf. fig. \ref{four_plots_diff_z}), absorbed AGNs with observed X-ray luminosities in the above range need to be more luminous than $\sim 10^{44.5}${} erg s$^{-1}${} intrinsically, and thus the rarity of such objects leads to the high fraction of unabsorbed sources in these bins. Secondly, as fig. \ref{zspec_HS}{} demonstrates, estimating column densities below N$_{H} = 10^{22}${} cm$^{-2}${} becomes increasingly harder for redshifts above z = 2.0. Thus, we caution that this increase in the fraction of absorbed AGNs with redshift in this luminosity regime may be caused by observational rather than truly evolutionary effects. \\
Correcting for the absorption in each source, leads to a considerable shrinking of the spread in the luminosities of the AGNs and disentangles the population of galaxies better from the faint AGN sources, as can be seen in fig. \ref{z_spec_vs_Lx_I}. Furthermore, even at low redshifts the absorbed and unabsorbed sources now cover largely overlapping parameter space. Indeed, fig. \ref{fraction_vs_zspec_II}{} shows that the ratio of absorbed to unabsorbed sources remains constant with redshift even at the high luminosity bin, if absorption corrections are taken before the binning. Despite the often large uncertainties of the corrections for individual sources, we believe this result to be robust in the statistical sense.\\

%\begin{figure}
%\resizebox{\hsize}{!}{\includegraphics[angle=270]{fraction_vs_luminosity.ps}}
%\caption[Luminosity vs. fraction]{The fraction of sources with log N$_H > 22.0${} for the AGN in the CDF-N versus their observed luminosity. The data points in black represent the complete sample, whereas the red,green and blue values are derived taking into account only sources within the indicated redshift bins. The dotted line represents the \citet[]{2003ApJ...598..886U} fit to their data.}
%\label{fraction_vs_luminosity_I}
%\end{figure}

%\begin{figure}
%\resizebox{\hsize}{!}{\includegraphics[angle=270]{fraction_vs_luminosity_corrected.ps}}
%\caption[Luminosity vs. fraction]{The fraction of sources with log N$_H > 22.0${} for the AGN in the CDF-N versus their absorption corrected luminosity. The data points in black represent the complete sample, whereas the red,green and blue values are derived taking into account only sources within the indicated redshift bins. The dotted line represents the \citet[]{2003ApJ...598..886U} fit to their data.}
%\label{fraction_vs_luminosity_II}
%\end{figure}
Is the absorption luminosity dependent ? Figure \ref{fraction_vs_zspec_II}{} shows that there is a significant decline in the fraction of absorbed sources towards higher observed luminosities. Furthermore, this trend remains the same for all redshifts, lending further credibility to the assumption of no redshift dependence of this ratio. Note that the decline in our sample is steeper than predicted by  the model of \citet[]{2003ApJ...598..886U}. If again we bin according to the absorption corrected luminosity, the trend becomes less prominent, but now exhibits the same slope as \citet[]{2003ApJ...598..886U}. Most importantly, however, for both cases the observed fraction of absorbed sources in our faint sample is higher than predicted by the \citet[]{2003ApJ...598..886U}{} model, especially in the low luminosity range, which is in agreement with the results from figs. \ref{obs_NH_distribution_different_types} and \ref{obs_NH_distribution_different_studies}. Thus, we propose that the absorption column density function $f = f(L_{X},z;N_{H}$) needs to be modified in order to produce more sources that are absorbed when going to fainter flux limits. While the drop in the fraction of absorbed sources with observed luminosity can be understood in the \citet[]{2004ApJ...616..123T}{} model by invoking the selection effects of redshift dependent absorption in the flux limited sample as described above, the decline of that fraction with absorption corrected luminosity indicates that the assumption of a luminosity independent torus geometry, structure and ionisation does not hold.\\
Relying on the simple hardness ratio method to derive the column density and neglecting spectral details, possibly introduces systematic effects that skew the resulting N$_H${} distribution. Ignoring a possible contribution to the soft X-ray spectrum caused scattering and/or reflection of the original nuclear source's spectrum or the extension of hot thermal emission by a starburst, or a narrow-line region, causes us to underestimate the absorption - and thus leads to a strenghtening of our argument : the effects of ignoring a soft-component are more severe for the lower redshift sources, and it is here where the main difference to \citet[]{2003ApJ...598..886U}{} arises. If the column density reaches values above N$_{H} = 10^{24}${} cm$^{-2}$, the AGN is totally buried ('Compton-thick') and possibly only a reflection component is seen which mimicks an unabsorbed source at much lower luminosity. \citet[]{1995ApJ...454...95M,2002AJ....124.1839B}, however, indicate that such sources do not make up a significant fraction of our AGN sample. The most important simplification in this respect is fixing of the value for the slope of the intrinsic power-law to $\Gamma = 1.8$, which is clearly an 'impossible' choice for some sources (cf. the data points below 'zero' absorption in fig. \ref{zspec_HS}). While we have tested our approach with a range of slopes (1.5-2.2) and all the major trends remained unchanged, and we are therefore convinced of their validity for the statistically large enough sample, we believe that a more detailed spectral fitting can certainly put these results on a more solid base. The depth of the surveys as of now, however, limits the ability to perform such analyses for most of the sources as the median number of photons for the sources with optically determined redshifts is only about 70. Any extensions of the Deep Fields towards 3-5 Ms would certainly allow for more detailed analyses and eleviate the need of focusing on the brightest members, as e.g. \citet[]{2004AJ....128.2048B}{} also point out.
%%%%%%%%%%%%%%%%%%%%%%%%%%%%%%%%%%%%%%%%%%%
\subsection{Fits to the NIR-optical-UV broadbands and N$_H${} estimates}
\begin{figure}
\resizebox{\hsize}{!}{\includegraphics[angle=270]{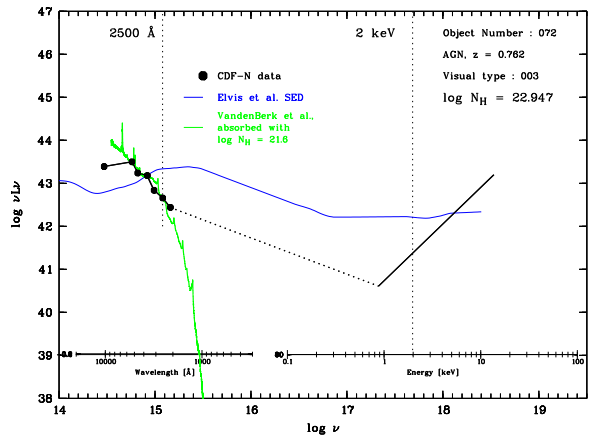}}
\caption[Fitting example]{Example of a fit to the broadband NIR-optical-UV SED of an AGN classified as having substantial absorption both in the X-ray and the optical wavelength regime. The spectrum of VandenBerk et al. (2005) is absorbed by dust producing an extinction following the Mathis (1990) measurement for the SMC and assuming an R$_{V}$=3.1, and the n folded with the appropriate filter functions for the available bands, applying the band correction terms of Capak et al. (2004). Note that the HK band only moves fully into the vandenBerk template at redshifts greater than z=1.805, sources with lower redshifts have thus been fitted without taking this band value into account. For some of the sources, the Elvis et al. (1994) template provided a better fit for the HK and z bands, probably due to the fact that the vandenBerk spectrum does contain some host galaxy contribution. The X-ray part of the spectrum is plotted as pure power-law with index $\Gamma$ taken from Alexander et al. (2003). The value of N$_{H}${} in the upper right corner refers to the measurement obtained from the X-ray analysis, and the object number is from the catalogue of Alexander et al. (2003). The corresponding graphs for each single source can be obtained in the electronic version of the journal.} 
\label{fitting_example}
\end{figure}

\begin{figure}
\resizebox{\hsize}{!}{\includegraphics[angle=270]{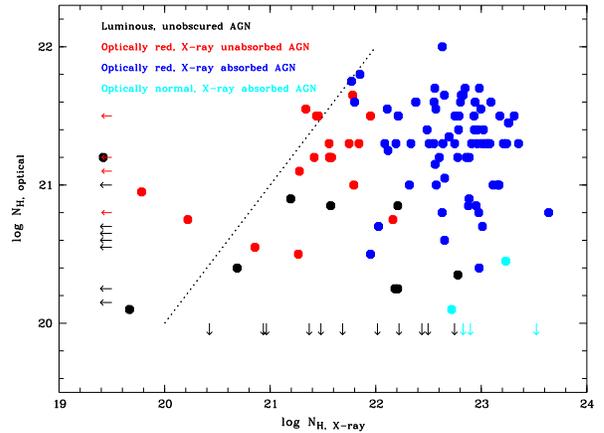}}
\caption[NH Comparison]{Comparison of the absorbing column densities derived by X-ray and optical analyses, N$_{H,X}${} and N$_{H,O}$ for the various types of AGNs classified in section 3. The dotted line indicates a ratio of unity. Objects for which either the X-ray or the optical column density could not be measured with the methods detailed in the text, arrows indicate upper limits. Note the absence of a correlation even in the case where both methods show substantial absorption (blue data points).} 
\label{NH_comparison}
\end{figure}

In order to determine whether the X-ray absorption is connected to absorption and reddening of the sources caused by dust, we have fitted the broadband SEDs of each AGN with a simple one-parameter model. Assuming a standard-gas-to dust ratio, R$_V${}=3.1, and the extinction curves derived by \citet[]{1990ARA&A..28...37M}{} for the SMC, we found the best-fitting value for N$_H${} by reddening the standard quasar spectrum of \citet[]{2006AJ....131...84V}. We then fold the resulting absorbed spectrum through the appropriate filter functions for the different instruments, and appl the band correction terms found by \citet[]{2004AJ....127..180C}{} in order to match the up to seven broadband fluxes for each AGN. An example for this fitting procedure is shown in figure \ref{fitting_example}.  
While this procedure relies upon a very simplistic picture of the AGN SED - leaving out any contribution of the host - it is able to fit the various types identified in the first sections remarkably well in most cases. Especially for the high luminosity and high redshift sources, we achieve an excellent match with an absorbed Vandenberk spectrum. In some cases, however, the NIR bands (namely the HK' and z band values) are overestimated by our approach. We attribute this to the possible inclusion of host galaxy contribution in the Vandenberk spectrum, and indeed, in those cases, the Elvis et al. standard SED, where efforts were made to specifically reduce the amount of host contribution, provides a better fit in these bands. We have grouped the AGNs according to the goodness of the fit, and as the table in the appendix shows, the majority of the bright sources fall in the highest confidence category, while the AGNs with apparent outliers are mainly the low luminosity, low redshift ones.\\
When comparing the column densities derived by the two independent methods in the X-ray and NIR-optical-UV bands for the various types of AGNs identified by our classification scheme (see fig. \ref{NH_comparison}), it becomes clear that there is no apparent correlation between the absorption measured by the two different methods. This is, of course, not surprising for the AGNs where we only see signs of absorption in one of the regimes probed. But even in the cases where we have a clear signature of absorption in both the X-rays and the optical, there is no relation between the two values of the column density, as fig.\ref{NH_comparison}{} clearly shows.\\
A variety of explanations can be brought forward to explain the lack of a correlation : firstly, the X-ray absorption, mainly by the metals in the gas phase of the absorber/absorbers, and the reddening by the dust in the NIR to the UV do not have to coincide spatially. In fact, there is evidence that the material responsible for both absorption processes is spread out over wide ranges of the parameter space (distance from the central source, temperature, density and composition). Secondly, our approach for measuring the column density by fitting the NIR-UV broadband relies on assumptions about the dust that not necessarily have to hold in the environments probed here - there are indications that the dust composition and thus the reddening laws around the central enginges of AGN are very different from the 'canonical' values used here : see e.g. \citet[]{2001A&A...365...28M, 2004A&A...420..889M}{} and references therein. In any case, it is clear that we are far from understanding the absorption processes that are responsible for dimming QSO light in various basspands. This is not only an interesting topic on its own merits, it is furthermore an extremely important step towards being able to characterise the AGN population as a whole and comparing results of deep surveys in different wavelength regimes. Unfortunately, with the data at hand it is not possible to pursue a more detailed study in this context, especially for the X-ray spectroscopy, as pointed out in the last section.      

%%%%%%%%%%%%%%%%%%%%%%%%%%%%%%%%%%%%%%%%%%%%%%%%%%%%%%%%%%
%%%%%%%%%%%%%%%%%%%%%%%%%%%%%%%%%%%%%%%%%%%%%%%%%%%%%%%%%%

\section{The UV and X-ray Luminosity relation}\label{UV-Xray}

\begin{figure}
\resizebox{\hsize}{!}{\includegraphics[angle=270]{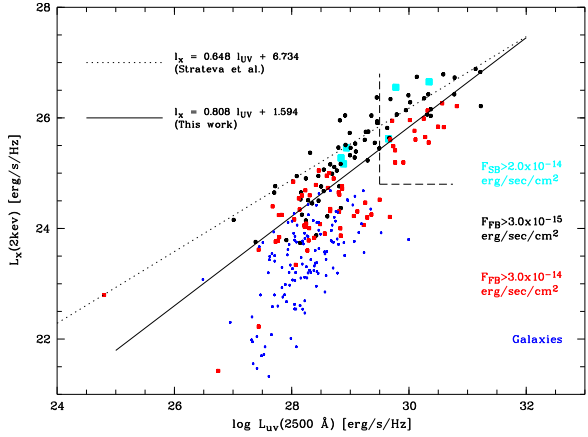}}
\caption[2keV vs. 2500 \AA{} monochromatic luminosities]{Monochromatic 2keV X-ray Luminosities versus the monochromatic UV luminosities at 2500 \AA{} for the CDF-N sources. The X-ray luminosity are calculated from the soft-band (0.5-2.0 keV) flux by assuming a power-law X-ray spectrum with the appropriate photon index $\Gamma _{eff}${} for each source. UV luminosities are inferred by fitting a 5th degree polynomial to the broadband SED in the 7 bands described in paragraph \ref{Data_analysis}. The dotted line represent the correlation for X-ray bright, optically selected AGNs as found by \citet{2005AJ....130..387S}. This fit is only a good result for the subsample of bright sources. If the faint sources are included, the slope for the correlation steepens and the scatter increases. Note that all faint sources are found to be below the \citet{2005AJ....130..387S}{} relation. Additionally, there is a significant overlap of the parameter space occupied by the galaxies and the faint AGNs. For objects more luminous than indicated by the dashed lines we expect the dominant contribution to the flux to be arising from the nuclear source rather than the host galaxy.}
\label{lx_luv_relation}
\end{figure}
For a comparison of X-ray surveys to those in the optical/UV, it is necessary to understand the relations between the emission in these two wavelength/energy regimes. \citet[]{1982ApJ...262L..17A}{} characterised an AGN sample by introducing $\alpha _{ox} = -0.3838$ log[F$_{\nu}$(2keV)/F$_{\nu}$(2500 \AA)], the slope of a hypothetical power law extending between 2500 \AA{} and 2 keV in the rest frame. It is well known that $\alpha _{ox}${} decreases with UV luminosity \citep[]{1990ApJ...360..396W}.\\
Recently, \citet{2005AJ....130..387S}{} and Steffen et al. (2006) have evaluated the dependence of $\alpha _{ox}${} on the source luminosity for a sample of 228 optically selected AGNs with ROSAT detections. \citet{2005AJ....130..387S}{} report a linearly decreasing $\alpha _{ox}(l _{UV})${} as a function of increasing source UV luminosity:
\begin{equation}\label{alpha_ox}
\alpha _{ox}(l _{UV}) = - (0.136 \pm 0.008)l _{UV} + (2.616\pm 0.249)
\end{equation}
for their complete sample including the high-z and Sy 1-galaxies.\\
We have computed the monochromatic X-ray luminosities at rest-frame 2 keV for the CDF-N sources from their soft-band (0.5-2 keV) flux by assuming a power-law X-ray spectrum with the appropriate photon index $\Gamma _{eff}${} for each source and the Galactic hydrogen column density for the CDF-N. Additionally, we have derived the flux density and hence monochromatic UV-luminosity at rest frame 2500 \AA{} by fitting a 5th degree polynomial to the broadband spectral energy distribution given by the up to 7 photometric measurements described in paragraph \ref{Data_analysis}. We note that these two methods are slightly different from the \citet{2005AJ....130..387S}{} approach where a fixed $\Gamma _{eff} = 2.0${} and $\alpha _{o} = -0.5${} for the power-law slopes are used to extrapolate to the rest-frame 2 keV resp. 2500 \AA. The differences in the derived quantities for the range of our parameters do not exceed 20 \%, however.\\%Furthermore, we did not correct for any host galaxy distribution in the broadband spectral energy distributions, which might lead to an overestimate for the UV luminosity especially for the low redshift, low luminosity sources as \citet[]{2005AJ....130..387S}{} point out.\\
Fig. \ref{lx_luv_relation}{} shows the X-ray luminosity at 2 keV vs. the UV luminosity at 2500 \AA{} for the CDF-N sources. The sources in blue are the AGNs that pass the X-ray flux limit for the \citet{2005AJ....130..387S}{} sample ($F _{0.5-2.0 keV} > 2 \times 10 ^{-14}${} erg s$^{-1}${} cm$^{-2}$), black dots represent the objects that meet the criterion to be included in the \citet{2003ApJ...598..886U}{} sample ($F _{0.5-8.0 keV} > 3 \times 10 ^{-15}${} erg s$^{-1}${} cm$^{-2}$), and the data points in red mark the galaxies in our sample as defined by \citet{2004ApJ...607..721N}. The dotted line represents the fit of \citet{2005AJ....130..387S}, which is in excellent agreement with the data in blue and in fairly good agreement with the data for the \citet{2003ApJ...598..886U}{} flux limit. If, however, we include the faint AGNs, we obtain a much steeper slope and larger scatter for the relation :
\begin{equation}\label{lx_luv_our_relation}
l _{X}(l _{UV}) = (0.808\pm 0.047)l _{UV} + (1.594\pm1.362)
\end{equation}
All of the data points for the faint sample actually lie under the \citet{2005AJ....130..387S}{} relation. Also note that there is a substantial overlap of the parameter space occupied by the 'Norman' galaxies and the AGNs.
\begin{figure}
\resizebox{\hsize}{!}{\includegraphics[angle=270]{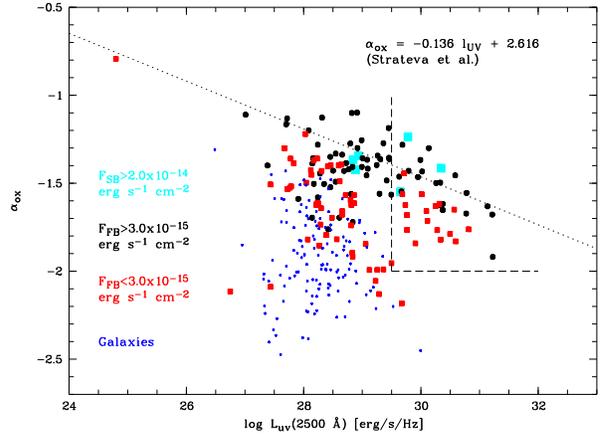}}
\caption[$\alpha _{ox}${} vs. 2500 \AA{} monochromatic luminosities]{The ratio of the monochromatic fluxes at 2 keV and 2500 \AA{}, $\alpha _{OX}$, versus the UV luminosity  of the CDF-N sources. The dotted line represents the anticorrelation for X-ray bright, optically selected AGNs as derived by \citet{2005AJ....130..387S}. It holds only for the subsample of bright sources, while the inclusion of the faint AGNs leads a situation with no clear dependence of $\alpha _{OX}${} with UV luminosity. Note again the overlap area for galaxies and the faint sources. For objects more luminous than indicated by the dashed lines we expect the dominant contribution to the flux to be arising from the nuclear source rather than the host galaxy.}
\label{aox_luv_relation}
\end{figure}
Fig. \ref{aox_luv_relation}{} shows $\alpha _{ox}${} vs. the monochromatic UV luminosity. Again, the \citet{2005AJ....130..387S}{} fit describes the data for the bright samples very well. The inclusion of the fainter sources, however, leaves us with no clear dependence of the X-ray-to-UV index with UV luminosity.\\
Why do the fainter sources fall into different parameter spaces ? We have investigated three possibilities:
\begin{enumerate}
\item The sources could be intrinsically X-ray underluminous compared to their brighter counterparts at the same UV luminosity.
\item There could be a higher X-ray absorption column density while the absorption in NIR/optical/UV stays the same. 
\item There could be a substantial contribution of the host galaxy for the UV flux leading to an overestimate of the UV luminosity of the sources. The contribution to the X-ray flux, however, would be negligible.
\end{enumerate}
While the first two explanations shift data points in fig. \ref{lx_luv_relation}{} downwards, the last one results in a horizontal shift to the right.\\% Interesting examples of  AGNs that appear to be truly X-ray underluminous are 3C351 \citep[]{1993ApJ...415..129F}{} and PHL1811 \citep[]{2001AJ....121.2889L}.\\
First, we have tested whether deviations from the broad-band optical/NIR SEDs of the faint sources compared to the bright sample and/or similarities with the SEDs of the galaxies, indicate that this wavelength regime is dominated by the host's stellar emission. We found that there is no difference in the colours for these three subsamples of the CDF-N : the regions occupied by bright and faint AGNs as well galaxies in the (B-V)-versus-(B-z) space largely overlap. Because the observed colours are, of course, quite redshift dependent, we have additionally compared only objects of the same redshift range. Even then, there is virtually no difference in the broad-band SED shapes for X-ray faint and bright AGNs. Thus, we rule out that a \emph{difference} in the host galaxy contribution to the UV luminosity estimate is the reason for the difference in $\alpha _{ox}${} for the two subsamples(i.e. hypothesis 3).\\
For absorbed AGNs, this result is not surprising : assuming an L$_*${} elliptical galaxy as host, a galactic gas-to-dust ratio and extinction curve for the nuclear source and a standard QSO SED, the contribution of an AGN of moderate X-ray luminosity (L$_{X} = 10^{43}$erg s$^{-1}$) and a N$_{H} = 10^{22}$cm$^{-2}${} to the optical emission is a marginal 4\%{} at z=0, as \citet{2004ApJ...616..123T}{} point out.  Most of the low redshift and low luminosity sources in both the bright and faint subsample have values similar to this example. Towards higher redshifts the AGN component becomes more important as more regions of the spectrum are sampled where stellar emission fades away rapidly. Therefore, we will focus in the following solely on the difference between the bright and faint sources in our sample that lie above L(2500 \AA) = $10^{29.5}$ erg s$^{-1}$Hz$^{-1}${} and L$_{X}$(2keV) = $10^{24.5}$ erg s$^{-1}$Hz$^{-1}$, as indicated by the dotted region in fig. \ref{lx_luv_relation}, where we expect the dominant contribution to the flux at 2500 \AA{} to be arising from the nuclear source.\\
The X-ray faint sources are in general at high redshifts (z$>$2.0) and exhibit observed rest-frame full band X-ray luminosities that are on average a factor of 6 lower than their X-ray bright counterparts. While their broadband optical colours are virtually the same, these X-ray faint AGNs tend to be heavily absorbed (cf. fig. \ref{strateva_highL_II}). All but one of the faint sources that have a hardness ratio that enables an absorption column density estimate with our model parameters, exhibit a log N$_H > 22.5$, while only three of the 22 of the bright subsample sources have these large column densities. Therefore, absorption in the X-rays is an important factor for the faint sources to deviate from the \citet[]{2005AJ....130..387S}{} relations that are based on luminous, X-ray bright objects (hypothesis 2 from above). 
However, even if corrections to the full band X-ray luminosities based upon the column density derived by the hardness ratio method are applied to  both subsamples, the faint sources fall short of their bright counterparts at the same redshifts : the mean luminosity for the faint sample is $log L_{X,unabs.} = 44.1 \pm 0.2$, whereas the mean luminosity for the bright sample is log $L_{X,unabs.} = 44.6 \pm 0.3$. Thus, we do not rule out that intrinsic 'under-luminosity' (compared to the bright sample) is a minor component contributing to the trend for faint sources to fall below the \citet[]{2005AJ....130..387S}{} anti-correlation. Fig.\ref{alpha_ox_lx}{} not only demonstrates this underluminosity, but in addition sheds some more light on the relation between the UV and X-ray luminosities and their relations. While even before applying corrections there is no trend for $\alpha_{OX}${} with X-ray luminosity discernible, the absorption corrected values present a clear scatterplot - confirming the notion of Steffen et al. (2006) who also attest that their is no correlation. The four sources with the lowest observed luminosity exhibit extremely large column densities (log N$_{H,X} >$ 23.0) and thus enormous correction factors to the 2keV flux density, despite appearing virtually unabsorbed and reddened in the optical. We caution, however, that the uncertainties in the X-ray derived column densities are rather large , as table \ref{appendix_table}{} shows. Three of the four sources (their numbers in the original Alexander et al. (2003 catalogue are 51, 171, 190 and 330, respectively) have more than 140 full-band counts, and thus a full spectral analysis is feasible for them. We will examine these interesting sources that show strong X-ray absorption and apparently no dust extinction in greater detail in an upcoming analysis.\\
Unfortunately, both the number of sources for which we can make reliable comparisons as well as their range in absorption corrected luminosities and redshifts is too small to arrive at a definite statement for the existence and importance of true X-ray underluminosity. We do point out, however, that there are several such examples in the local universe : 3C351 \citep[]{1993ApJ...415..129F}{}  and PHL1811 \citep[]{2001AJ....121.2889L}{} fall well short of the expected X-ray luminosity when extrapolating from their UV fluxes.   

\begin{figure}
\resizebox{\hsize}{!}{\includegraphics[angle=270]{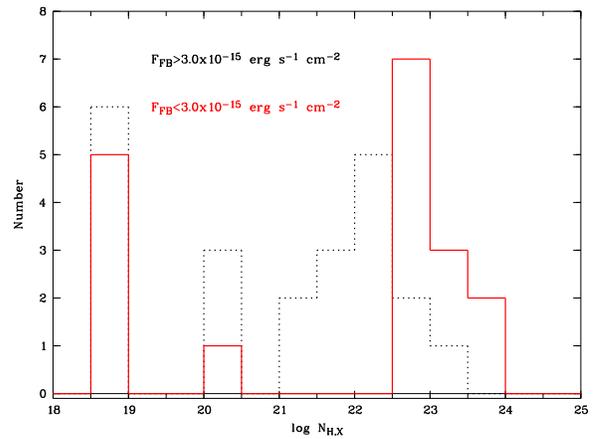}}
\caption[Colour_vs_log NH]{The column density distribution of AGNs more luminous than  L(2500 \AA) = $10^{29.5}$ erg s$^{-1}$Hz$^{-1}${} and L$_{X}$(2keV) = $10^{24.5}$ erg s$^{-1}$Hz$^{-1}$, as indicated by the dotted region in fig. \ref{lx_luv_relation}. Sources in red represent the faint sample, while data in black are from the bright sample. All but one of the X-ray faint sources exhibit signs of high absorption when the column density can be reliably estimated, whereas the bright objects tend to be un- or only moderately absorbed.}
\label{strateva_highL_II}
\end{figure}

\begin{figure}
\resizebox{\hsize}{!}{\includegraphics[angle=270]{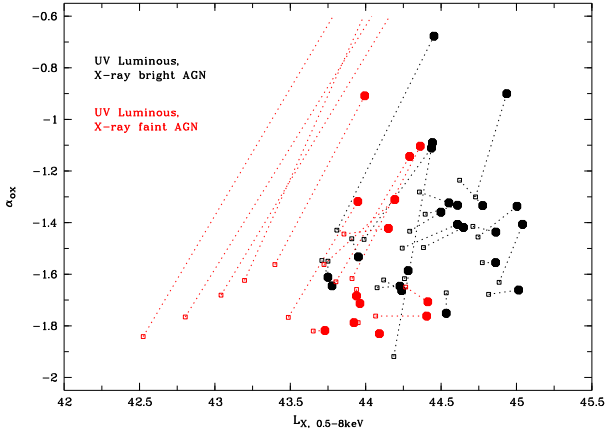}}
\caption[alpha_ox_vs_lx]{Observed and absorption corrected full band X-ray luminosities and $\alpha_{ox}$ values for the UV-luminous subsample within the dotted region of fig.\ref{lx_luv_relation}. The data points in black are for the bright sample (f$_{X} > 3 \times 10^{-15}$erg s$^{-1}$), whereas the red points are for the sample fainter than this limit used by Ueda et al. The open symbols represent the observed values for both the X-ray full band luminosity and $\alpha_{ox}${}, whereas the large, filled symbols are derived after correcting for absorption in both the X-ray and the UV regime. Note the absence of a relation between L$_{X}${} and $\alpha_{OX}$. Furthermore, even after applying the absorption correction the X-ray faint subsample resides at a lower luminosity than the X-ray bright counterparts albeit having the same UV luminosity distribution (cf. fig.\ref{lx_luv_relation}). The four sources with the lowest X-ray luminosity exhibit the most dramatic, yet uncertain correction for $\alpha_{OX}$. For further explanation of their nature cf. text.}
\label{alpha_ox_lx}
\end{figure}

%%%%%%%%%%%%%%%%%%%%%%%%%%%%%%%%%%%%%%%%%%%%%%%%%%%%%%%%
%%%%%%%%%%%%%%%%%%%%%%%%%%%%%%%%%%%%%%%%%%%%%%%%%%%%%%%%

\section{Discussion and Conclusions}\label{Discussion}
We have analysed data from the two deepest X-ray surveys to date, the Chandra Deep Fields North and South. By shifting the 270 sources of the CDF-N with spectroscopically secure redshifts into the emitted frame and visualising their broadband spectral energy distributions in a log$\nu L_{\nu }$-log$\nu${} plot, we were able to group the objects into five different classes.
\begin{itemize}
\item The low X-ray luminosity and low redshift bins are dominated by (numerically) by what appear to be normal starforming or starburst galaxies, consistent with the results of earlier studies (cf. e.g. \citet[]{2003AN....324...12H}).
\item The high redshift regime contains primarily unabsorbed, luminous AGNs that resemble the \citet{1994ApJS...95....1E}{} standard AGN.
\item At redshifts between $\sim$0.7-1.5{} there are numerous AGNs that show clear evidence for dust reddening and absorption in the optical/UV part of the broadband SED. About half of them, however, do not show any signs of significant X-ray obscuration, whereas the other half portray a $\Gamma _{eff} <<$1.9, indicative of column densities N$_{H}> 10 ^{22}$ cm$^{-2}$.
\item A small number of the AGN appear to be optically normal while exhibiting signs of strong X-ray absorption. These AGNs are of high UV luminosity (L$_{2500 \AA} > 10^{29.5}$erg s$^{-1}$Hz$^{-1}${} and have redshifts z$_{spec} > 1.6$. 
\item Based upon the hardness ratio, we have estimated the X-ray absorbing column density distribution of the spectroscopically surveyed CDF-N sources and the objects in the CDF-S with photometric redshifts. In order to assess the bias introduced by errors in the count-rates and the effects of having only few objects in the higher redshift bins, we performed a Monte-Carlo-analysis, creating 10000 mock data sets with the number count rate ratio fluctuating with the appropriate error distribution. While the spread in the N$_H${} estimate for each individual source can become rather large, the overall density distribution of the complete AGN sample remains robust against these randomisations, indicating the usefulness of the simple method in a statistical sense. The column density distribution for the sources classified as galaxies by \citet{2004ApJ...607..721N}{} peaks at much lower values than the one for AGN.
\item Including the faintest sources of the CDF-N (F$_{0.5-8.0 keV} < 3\times 10^{-15}${} erg s$^{-1}${} cm$^{-2}$) into the analysis, results in shifting the peak of the AGN observed N$_{H}${} distribution to higher values compared to the distribution that \citet{2003ApJ...598..886U}{} used. The faintest sources, which by and large are also less luminous than the sources included into the \citet{2003ApJ...598..886U}{} sample, thus exhibit signatures indicative of more X-ray absorption.
\item The strong redshift evolution seen in the ratio of unabsorbed to absorbed X-ray AGN (with the dividing line at N$_{H} = 10^{22}${} cm$^{-2}$) appears to be a selection effect. The cut-off $N_{H}${} value, beyond which the number of observed sources rapidly dwindles, coincides up to redshifts of $z\sim$1.5 very well with the $N_{H}${} value for which about 90\%{} of the soft-band flux is absorbed. This confirms the assumptions of \citet[]{2004ApJ...616..123T}{} and \citet[]{2003ApJ...598..886U}, but is in contrast to the findings of \citet[]{2005AJ....129..578B},who report a decreasing fraction of Type 1 AGN with redshift (their fig. 19), and the analysis of \citet[]{2006ApJ...639..740B}{}
who derive an evolving fraction of Type 2 AGN as $R \sim (1+z)^{0.3}$. However, we note that the classification scheme used in the latter studies is based on spectroscopic optical AGN features, whereas the former rely on an X-ray estimate of the absorbing column density.
\item The fraction of absorbed sources, however, decreases with luminosity : while 85($\pm$25)\%{} of the AGN with an observed full band, restframe X-ray luminosity of 41.5$<${} log L$_{X} <$43.0 have a column density above N$_{H} = 10^{22}$cm$^{-2}$, only 25($\pm$13)\%{} with 44.5$<${} log L$_{X} <$45.0 exhibit such high column densities. The tendency of this decrease agrees with the \citet[]{2003ApJ...598..886U}{} result, although we obtain a much higher fraction of absorbed sources towards lower observed X-ray luminosities. Thus, we propose that the absorption column density distribution function $f = f(L_{X},z;N_{H}$){} needs to be modified in order to produce the correct number counts of absorbed sources towards fainter fluxes.
\item Even after correcting luminosities for the effects of absorption, the decrease in the fraction of absorbed sources towards higher intrinsic luminosities remains statistically significant. This result contradicts the model of \citet[]{2004ApJ...616..123T}{} who assume a geometry and composition for the dust torus obscuring the central AGN that neither changes with redshift nor with luminosity.     
\item A comparison between the effective hydrogen column densities derived by the X-ray hardness ratio and by fitting the broadband optical continuum of the AGN reveals that there is no correlation between the two quantities, even in the cases where there are clear signs for absorption in both energy regimes. While this lack of a relation is significant and important to understand e.g. for the modelling of luminosity functions and their evolutions, it cannot be determined from the data available to us whether it is due to the different spatial extent of the absorbers and/or the composition of the dust surrounding the central engines.  
\item Focusing on the bright X-ray sources, we can reproduce the results of \citet{2005AJ....130..387S}{} who derived correlations between the monochromatic UV and the X-ray luminosities of a complete sample of optically selected SDSS AGNs with additional ROSAT detections. Including the faint X-ray sources, however, drastically steepens the slope and increases the scatter in the log $L_{X}$ - log $L_{UV}${} relation.
\item All of the X-ray faint sources (i.e. f$_{X, 0.5-8.0keV} < 3.0 \times 10^{-15}${} erg s$^{-1}$ cm$^{-2}$) fall below the \citet{2005AJ....130..387S}{} relation. We have investigated three possible scenarios to explain this behaviour. The sources could either be intrinsically X-ray underluminous compared to their X-ray bright counterparts at the same UV luminosity. They could exhibit a higher X-ray absorbing column density $N_{H}${} while remaining at the same level of optical and UV absorption. Or their UV luminosities might be overestimated because of a significant host galaxy contribution in this band, while the X-ray band remains virtually undiluted. While we cannot rule out the latter for the least luminous sources, we see no differences in the shapes of the NIR/optical/UV SED for those two subsamples of sources. Thus, we deem it unlikely that differences in the host dilution play an important role in this case. Focusing on the X-ray and UV luminous AGNs, we conclude that sources of fainter X-ray flux are more X-ray absorbed. However, even when correcting for this absorption they remain by a factor of $\sim${} 3 less luminous than their bright counterparts at the same redshifts. The subsample of such objects, which allow for a reliable estimate of the absorber column density, does not cover enough redshift and luminosity space to confirm this hypothesis.  
\item The $\alpha _{ox}-l_{UV}${} anti-correlation as seen by \citet{2005AJ....130..387S}{} also only holds for the X-ray bright sources. The inclusion of the fainter objects leads to a disappearance of any clear relation between $\alpha _{OX}${} and $L _{UV}$. Again, the higher absorption in the faint sources is the dominant factor in explaining the drop for the faint sources, with a possible small contribution by their lower luminosity (compared to the X-ray bright sources at similar redshifts).
\item Furthermore, whereas the bright X-ray sources could be in general easily distinguished from objects that appear to be galaxies, the fainter X-ray sources occupy regions in a multitude of parameter spaces that largely overlap with the areas held by the galaxies. Potentially a large number of the 'Norman' sources might harbour AGN activity - and/or that the host galaxy contribution to the observed UV luminosity of some sources classified as AGNs might be substantial. This agrees with the analysis of \citet[]{2004AJ....128.2048B}{} who derive quite different AGNs and 'galaxy' samples for the same underlying population.
\item Incorporating the sources of the CDF-S 1Msec catalogs strengthens all of our results above, as the redshift, luminosity and column density distributions are very similar to the CDF-N objects. However, since we had to rely on photometric rather than spectroscopic redshifts for the majority of these sources, we decided to focus on the CDF-N catalogs first and present the results without mingling the two samples.       
\end{itemize}  
From this analysis of the two Chandra Deep Fields, it is obvious that a more detailed study of the effects of absorption in both the X-rays and the NIR/optical/UV wavelength regime is crucial to working out the evolution of the underlying source population. In addition, the effects of the possible contribution of host galaxy light to the optical/UV emission for the fainter X-ray sources classified as AGNs, need to be addressed thoroughly. On the other hand, many of the sources assumed to be star-forming or starburst galaxies, can in fact harbour AGNs. We need to find ways to properly disentangle the stellar emission from the AGN light output. Currently, we are working on implementing methods similar to the ones used e.g. by \citet[]{2006AJ....131...84V}{} for decomposition of the stellar and AGN contribution to the SED.\\
%%%%%%%%%%%%%%%%%%%%%%%%%%%%%%%%%%%%%%%%%%%%%%%%%%%%%%%%%%
%%%%%%%%%%%%%%%%%%%%%%%%%%%%%%%%%%%%%%%%%%%%%%%%%%%%%%%%%%
%\begin{acknowledgements}
%We would like to thank .......
%comments on the manuscript.
%\end{acknowledgements}
\bibliographystyle{aa}
\bibliography{sfrank}

%%%%%%%%%%%%%%%%%%%%%%%%%%%%%%%%%%%%%%%%%%%%%%%%%%%%%%%%%%
%%%%%%%%%%%%%%%%%%%%%%%%%%%%%%%%%%%%%%%%%%%%%%%%%%%%%%%%%%
\section{Appendix A : The Properties of the 2Ms CDF-N X-ray objects with spectroscopic redshifts}
The following table contains the properties derived for all of the 270 CDF-N sources with confirmed spectroscopic redshifts. This table is available in its entirety in the electronic version of the (magazine to publish our paper here). A portion is shown here for guidance regarding its form and content.\\
The catalogue number refers to the original listing of \citet{2003AJ....126..539A}.\\
\clearpage
%% Remove the two lines and the last line if you want
%% want to incorporate this table into another LaTex document.

%% The values (usually only l,r and c) in the last part of
%% \begin{deluxetable}{} command tell LaTeX how many columns
%% there are and how to align them.
\begin{table}

%% Rotate to a landscape orientation
%\rotate

%% Over-ride the default font size
%% Use 8pt
%\tabletypesize{\scriptsize}

%% Use \tablewidth{?pt} to over-ride the default table width.
%% If you are unhappy with the default look at the end of the
%% *.log file to see what the default was set at before adjusting
%% this value.

%% This is the title of the table.
\caption{Properties of the 2Ms CDF-N X-ray objects with spectroscopic redshifts.}\label{appendix_table}
\scriptsize
\begin{center}
\begin{tabular}{|c|c|c|c|c|c|c|}
\hline
Catalog number & RA & DEC & z$_{spec}$(1) & Norman type (2) & Visual SED type (3) & \\
  & (Degrees) & (Degrees) &  &  & & \\
\hline
3 & 188.828415 & 62.26431 & 0.1380 & 0 & 3 & \\
5 & 188.838837 & 62.27447 & 0.5590 & 1 & 2 & \\
6 & 188.839996 & 62.30200 & 0.1350 & 0 & 2 & \\
11 & 188.872421 & 62.21567 & 2.4130 & 1 & 3 & \\
16 & 188.904587 & 62.28989 & 2.0500 & 1 & 1 & \\
\dots & \dots         & \dots       & \dots     & \dots & \dots & \\

\hline
\hline
Catalog number & log L$_{X}$(0.5-8.0 keV) & HR (4)& log N$_H, X$ (5) & 1$\sigma${} error on log N$_H$ & log L$_{X, abs. corr.}$ (6) & log L$_{2500 \AA}$ (7) \\
 & (erg/sec) &  & (cm$^2$) & (cm$^2$) & (erg/s) & (erg/s) \\
\hline
3 & 40.85 & 0.250 & -1.000 & -1.000 & 41.057 & 27.911 \\
5 & 42.85 & 0.399 & 21.562 & 0.4004 & 43.021 & 28.502 \\
6 & 41.08 & 0.667 & -1.000 & -1.000 & 41.250  & 27.833 \\
11 & 44.24 & 0.351 & 22.573 & 0.1840 & 44.608 & 30.258 \\
16 & 44.11 & 0.493 & 20.687 & 1.7640 & 44.238 & 30.311 \\
\dots & \dots    & \dots    & \dots     & \dots     & \dots     & \dots     \\
\hline
\hline
Catalog number & $\alpha_{OX}$ & log L$_{HK}$ (8) & log L$_{z}$ (8) & log L$_{I}$ (8) & log L$_{R}$ (8) & log L$_{V}$ (8) \\
 & & (erg/s) & (erg/s) & (erg/s) & (erg/s) & (erg/s) \\
\hline
3 & -2.00 & 43.259 & 43.691 & 43.471 & 43.566 & 43.431 \\
5 & -1.42 & 43.580 & 43.812 & 43.592 & 43.688 & 43.512 \\
6 & -1.88 & 43.398 & 43.630 & 43.450 & 43.466 & 43.330 \\
11 & -1.49 & 45.076 & 45.028 & 44.727 & 44.783 & 44.607 \\
16 & -1.62 & 44.822 & 44.893 & 44.793 & 44.889 & 44.953 \\
\dots & \dots    & \dots     & \dots     & \dots     & \dots     & \dots     \\
\hline
\hline
Catalog number & log L$_{B}$ (8) & log L$_{U}$ (8) & FB X-ray counts & log N$_{H, opt.}${} (9) & Goodness of NIR-UV fit  & Comments \\ 
 & (erg/s) & (erg/s) &  & (cm$^2$) & (10) & (11)  \\ 
\hline
3 & 43.124 & 42.703 & 112.8 & \dots & \dots & \dots  \\ 
5 & 43.486 & 43.345 & 596.8 & 21.20 & 1 & HK$_{low}$ \\
6 & 43.104 & 42.763 & 216.6 & \dots & \dots & \dots \\
11 & 44.701 & 44.120 & 462.6 & 21.00 & 2 & VB \\
16 & 45.087 & 45.346 & 512.3 & 20.40 & 1 & VB,U$_{high}$,B$_{high}$ \\
\dots & \dots     & \dots     & \dots     & \dots     & \dots & \dots \\
%% All data must appear between the \startdata and \enddata commands
%\startdata
\hline
\hline
%16 & 188.904587 & 62.28989 & 2.0500 & 1 & 1 & 44.11 & 0.493 & 20.687 & 1.7640 & 44.238 & 30.311 & -1.62 & 44.822 & 44.893 & 44.793 & 44.889 & 44.953 & 45.087 & 45.346 & 512.30 & 20.400 & 1 & VB,U_high,B_high \\
%.. & .. & .. & .. & .. & .. & .. & .. & .. & .. & .. & .. & .. & .. & .. & .. & .. & .. & .. & .. & .. & .. & .. & .. \\
%\hline
%\enddata
\end{tabular}
%% Include any \tablenotetext{key}{text}, \tablerefs{ref list},
%% or \tablecomments{text} between the \enddata and 
%% \end{deluxetable} commands

%% General table comment marker
%\tablecomments{Usual statement about : The full version can be obtained here.}

%% No \tablerefs indicated
\end{center}
{\bf Note (1):} Spectroscopic redshift from \citet[]{2003AJ....126..632B}\\
{\bf Note (2):} Type according to Norman et al. (2004). [0 = Galaxy, 1 = AGN]\\
{\bf Note (3):} Type according to the visual classification of this paper :\\
          1 = luminous, unobscured AGN\\
          2 = red \& X-ray flat AGN\\
          3 = red \& X-ray steep AGN\\
          4 = optically normal, X-ray steep AGN\\
          X-ray steep as defined in the log $\nu${} - log $\nu$f$_{\nu}${} plot (cf. text).\\
{\bf Note (4):} Hardness ratio defined as (H-S)/(H+S), where H and S refer to the 
          counts in the hard and soft bands.\\
{\bf Note (5):} Result of Monte-Carlo simulation. Average of 10,000 randomisations of the hardness ratios for the complete data set. The quoted 1-sigma error is derived from the sample statistics. A value of 18.99 indicates that the hardness ratio method does not provide a meaningful result. Sources classified as galaxies are denoted with a column density of -1.0.\\
{\bf Note (6):} Rest-frame and absorption correction using the column density from (5) and an unabsorbed power-law spectrum with photon index $\Gamma$=1.8.\\
{\bf Note (7):} Flux density at 2500 \AA derived by fitting 5th degree polynomial to NIR-UV SED.\\
{\bf Note (8):} Derived from measured HK, z, I, R, V, B and U brightnesses from the catalogue of Barger et al. (2003), corrected to AB magnitudes with correction terms by Capak et al. (2004).\\
{\bf Note (9):} The NIR-UV SED is fitted by absorbing the vandenBerk SDSS average AGN spectrum with an SMC extinction curve using the relation of \citet[]{1990ARA&A..28...37M} to convert N$_H${} into an extinction A$_{Ic}$. The resulting spectrum is folded through the appropriate filter functions of Capak et al.(2004), and the colours are normalised to the R band. A value of 19.00 indicates that the source shows no absorption.\\
{\bf Note (10):} Goodness of the fit in the NIR-UV :\\
               0 = unsatisfactory fit, at least three fit flux values deviate more than 3$\sigma${} from the measured flux\\
               1 = medium quality, 2 fitted flux values deviate more than 2$\sigma${} from the measured flux\\
               2 = good fit, only one fitted flux value is allowed to deviate more than 2$\sigma${} from the measured flux\\
           Note : The HK band only moves fully into the vandenBerk template at z=1.805. Lower redshift sources were therefore fitted without taking this band into account.\\
{\bf Note (11):} Comments on the shape of the SED and other characteristics.\\
              VB = SED fitted well by absorbed VandenBerk SED only.\\
              E  = SED fitted well by absorbed Elvis et al. SED only.\\
              BR = unusual break in the SED\\
              H  = SED shape hints at host contamination\\
              MW = Milky-Way type extinction curve possible (i.e. more absorption around 2500 \AA)\\
              \#$_{low}$/\#$_{high}${} = unusually low or high flux in band \#{} (compared to fit)\\
              no $\#${} = no reliable flux measurement in band \#{} available
\end{table}

\end{document}